\documentclass[%
 reprint,
 amsmath,amssymb,
prb,
]{revtex4-2}

\usepackage{graphicx}
\usepackage{dcolumn}
\usepackage{bm}
\usepackage{color}
\usepackage[colorlinks,
            linkcolor=blue,    
            anchorcolor=blue,  
            citecolor=blue,        
            ]{hyperref}

\begin{document}

\preprint{APS/123-QED}

\title{In-Plane Magnon Valve Effect in Magnetic Insulator/Heavy Metal/ Magnetic Insulator Device}

\author{Tianyi Zhang$^{1}$, Caihua Wan$^{1,\ 2*}$ and Xiufeng Han$^{1,2,3}$}

 \email{xfhan@iphy.ac.cn; wancaihua@iphy.ac.cn}
\affiliation{%
 ${}^{1}$Beijing National Laboratory for Condensed Matter Physics, Institute of Physics, Chinese Academy of Sciences, Beijing 100190, China\\
 ${}^{2}$Center of Materials Science and Optoelectronics Engineering, University of Chinese Academy of Sciences, Beijing 100049, China\\
 ${}^{3}$Songshan Lake Materials Laboratory, Dongguan, Guangdong 523808, China
}%

\date{\today}

\begin{abstract}
We propose an in-plane magnon valve (MV), a sandwich structure composed of ferromagnetic insulator/heavy metal/ferromagnetic insulator (MI/HM/MI). When the magnetizations of the two MI layers are parallel, the longitudinal conductance in the HM layer is greater than that in the antiparallel state according to the magnetic proximity effect, termed as the in-plane magnon valve effect. We investigate the dependence of MV ratio (MVR), which is the relative change in longitudinal conductance between the parallel and antiparallel MV states, on the difference in electronic structure between magnetized and non-magnetized metal atoms, revealing that MVR can reach 100\%. Additionally, the dependence of MVR on the thickness of metal layer is analyzed, revealing an exponential decrease with increasing thickness. Then we investigate the dependence of HM layer conductance on the relative angle between the magnetizations of two MI layers, illustrating the potential of MV as a magneto-sensitive magnonic sensor. We also investigate the effect of Joule heating on the measurement signal based on the spin Seebeck effect. Two designed configurations are proposed according to whether the electron current is parallel or perpendicular to the magnetization of the MI layer. In the parallel configuration, the transverse voltage differs between the parallel and antiparallel MV states. While in the perpendicular configuration, the longitudinal resistance differs. Quantitative numerical results indicate the feasibility of detecting a voltage signal using the first configuration in experiments. Our work contributes valuable insights for the design, development and integration of magnon devices.
\end{abstract}

\maketitle


Magnon valve (MV) is a sandwich structure consisting of ferromagnetic insulator/heavy metal/ferromagnetic insulator (MI/HM/MI), which provides a powerful platform for controlling the transport of magnon current \cite{3,2,1}. Similar to spin valve \cite{4,5}, which modulates resistance by manipulating the relative magnetization of the magnetic layers, MV can also modulate output magnon current by manipulating the relative magnetization of the magnetic layers. The parallel-state of MV facilitates magnon transmission while its antiparallel-state impedes it, which is called MV effect (MVE) \cite{3}. Due to the advantages of magnon, such as the absence of Joule heat during transport and the prospective development in microwave devices within the GHz-THz frequency band \cite{6,7,8,9,10,11,12,13,14,15,16}, MV has potential to serve as foundational components in the future information technology. Investigation of magnon transport in MV is pivotal for further practical applications of magnon devices. Despite numerous studies on MV manipulation and the fabrication of various structures \cite{1,2,3,17,20,19,18}, limitations of transport damping and low conversion ratios constrain their utility.

In experiment, the MV serves to control the activation and deactivation of magnon channels, which necessitates real-time detection of the magnetizations of two MI layers. Typically, two methods are employed to achieve this goal. The first method is to use a magnetic measurement instrument to measure the average magnetization of the MV, such as vibration magnetometer method, which is not compatible with integrated processes. The second method is to deposit a layer of HM
onto the MV to detect transmitted magnon currents using the inverse spin Hall effect (ISHE) \cite{3}. However, due to the limited conversion efficiency from spin current to electron current, the ISHE signals measured experimentally for both antiparallel and parallel MV typically exhibit magnitudes on the order of microvolts \cite{3}. The observed signal magnitude is very small, and the measured signal is also susceptible to ambient temperature, which hinder its practical application. Therefore, it is necessary to develop an approach that meets the requirements of integration, has high detection efficiency, robust to environment and enables all-electrical measurement of the relative orientation of magnetizations of MI layers in the MV under external magnetic field.

In this work, we propose a method to determining the relative orientation of magnetizations in two MI layers by measuring the conductance of the middle HM layer in MV. When the magnetizations of the two MI layers are parallel, the longitudinal conductance in the HM layer is greater than that in the antiparallel state due to the magnetic proximity effect (MPE). We analyze the dependence of MV ratio ($MVR^{\uparrow \uparrow, \uparrow \downarrow} \equiv\left(\sigma_{x x, \uparrow \uparrow}-\sigma_{x x, \uparrow \downarrow}\right) / \sigma_{x x, \uparrow \uparrow}$, where $\sigma_{x x, \uparrow \uparrow}$ and $\sigma_{x x, \uparrow \downarrow}$ are longitudinal conductance of metal layer in parallel and antiparallel MV states, respectively) on the difference in electronic structure between magnetized and non-magnetized metal layer atoms. Additionally, we investigate the relationship between $MVR^{\uparrow \uparrow, \uparrow \downarrow}$ and the thickness of the metal layer. Then we investigate the influence of Joule heating on the measurement signal based on the spin Seebeck effect (SSE).

The schematic illustrating the in-plane MV effect (IMVE) based on the MPE is presented in Fig. 1. The MV consists of three components: the upper and lower MI layers are composed of yttrium iron garnet (YIG) with in-plane magnetization, and the central HM layer is composed of platinum (Pt).

\begin{figure}[htbp]
\includegraphics[width = \linewidth]{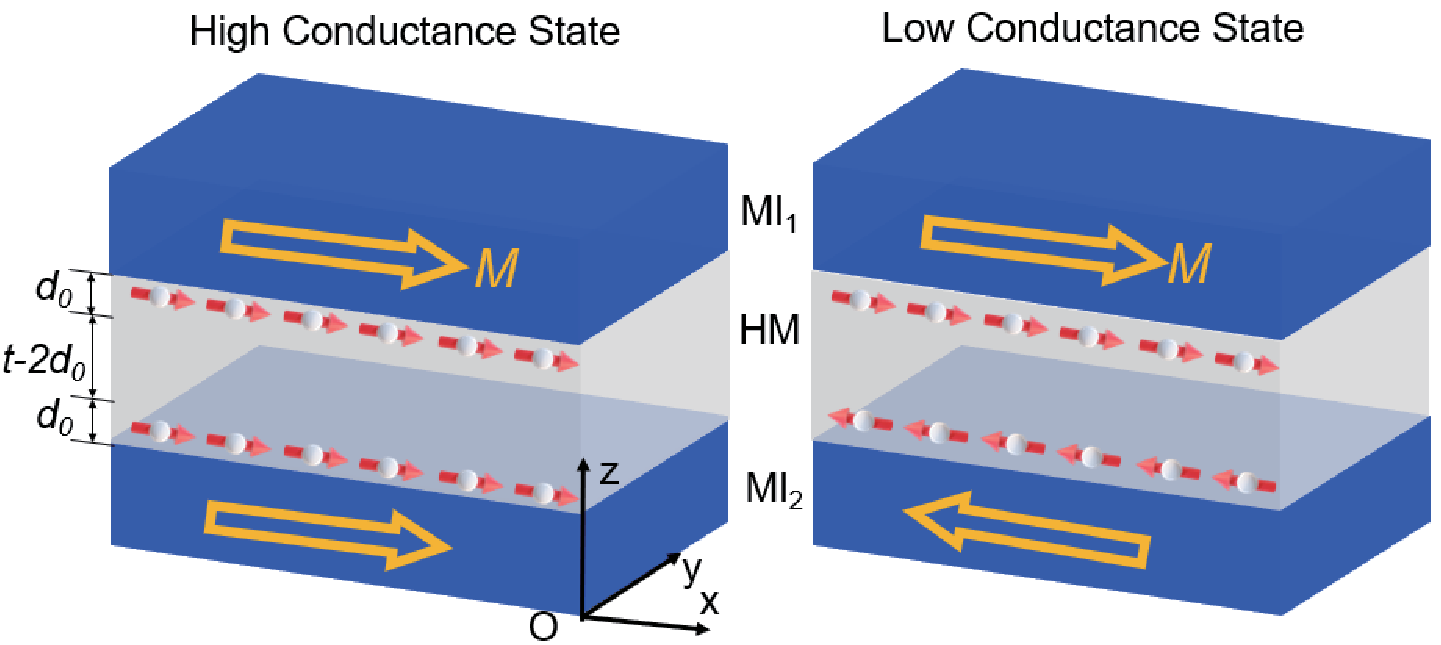}
\caption{Schematic of the IMVE based on the MPE. $t$ is the thickness of Pt layer, $d_0$ is characteristic length of the MPE.}
\end{figure}

Due to the MPE \cite{21,22,23}, the electronic structures of Pt atoms in the vicinity of the YIG interfaces are modified. Some Pt atoms undergo a transition from paramagnetic to ferromagnetic states, aligning their magnetic moments with those of the adjacent YIG. 
We assume the average magnetic moment of Pt atom $\mu_{Pt}(d)=\mu_0e^{-d/d_0}$ \cite{21,22}, where $\mu_0$ is magnetic moment of Pt atoms at YIG/Pt interface, $d$ is the distance from the Pt atom to the nearest YIG layer interface and $d_0$ is characteristic length of the MPE. There are two factors which cause $\mu_{Pt}$ to be negatively correlated with $d$: firstly, the proportion of magnetized Pt atoms decreases with increasing $d$; secondly, the magnetic moments of individual magnetized Pt atoms also decrease with increasing $d$. So far, neither experimental nor first-principles calculations have provided the contribution proportion of these two components. Here, we make a simplified assumption that both factors equally influence the decay of the average magnetic moment of Pt atoms.
Although this assumption may slightly deviate from reality, it does not affect the qualitative results. In our calculation, we consider the magnetic moments of unmagnetized and magnetized Pt atoms as follows: $\mu_{Pt}^{unm}=0,\ \ \mu_{Pt}^m(d)=\mu_0e^{-d/{2d}_0}=0.2\mu_Be^{-d/{2d}_0}$, where $\mu_B$ is Bohr magneton, $d_0=0.4\ $ nm \cite{21}. The proportion of magnetized Pt atoms is $n_{Pt}^m(d)/n_0=e^{-d/{2d}_0}$, where $n_0$ is the number of magnetized Pt atoms at the YIG/Pt interface. When the thickness of Pt $t>2d_0$, we can divide the Pt layer into three parts: two magnetization layers at the top and bottom, and an unmagnetized layer in between. Therefore, the Pt layer is analog to a spin valve with a ferromagnet/metal/ferromagnet sandwich configuration. The alteration in YIG magnetic state leads to a variation in the resistance of Pt layer. In the parallel state, the Pt layer exhibits high conductance, while in the antiparallel state, the Pt layer exhibits low conductance.

Here we use the coherent potential approximation (CPA) \cite{24,25,26} to calculate the conductance of the Pt layer. The CPA is a well-established method for calculating the conductance of binary alloys \cite{25,26}. We treat the magnetized Pt atoms and non-magnetized Pt atoms as two distinct atoms coexisting in the Pt layer.

Using the tight binding approximation model, the Hamiltonian of the Pt layer electronic system has the form

\begin{equation}\label{eq1}
\begin{array}{ll}
    \hat{H}&=\sum_{s,i}{\varepsilon_{is}\hat{a}_{is}^\dag \hat{a}_{is}}+\sum_{s,<i,j>}{t_{ijs}\hat{a}_{is}^\dag \hat{a}_{js}}
\end{array} 
\end{equation}
Where $\hat{a}_{is}^\dag(\hat{a}_{is})$ is the creation (annihilation) operator for electrons with spin s at the i-th layer, $\varepsilon_{is}$ and $t_{ijs}$ are on-site energy and nearest transition energy, 
$<i,j>$ denotes sum over the nearest lattice point.

The on-site energy at the n-th layer $\varepsilon_{ns}$  takes values $\varepsilon^m_{ns}$ and $\varepsilon^{unm}_{ns}$ with probabilities $x_n$ and $y_n=1-x_n$ respectively, where $x_n$ is the proportion of magnetized Pt atoms at the n-th layer, $t_{ijs}=t$ is the transition energy. Using the coherent potential, we can calculate the conductance of Pt \cite{24,25,27}

\begin{equation}\label{eq2}
\sigma_{x x}^{M P E} \propto \sum_{s, m, n} \frac{\delta_{n m}+\left(1-\delta_{n m}\right) \times\left(\frac{\left(\Delta_{n s}+\Delta_{m s}\right)^{2}}{\left(\Lambda_{n s}-\Lambda_{m s}\right)^{2}+\left(\Delta_{n s}+\Delta_{m s}\right)^{2}}\right)}{\left(\Delta_{n s}+\Delta_{m s}\right)}
\end{equation}

where $\Lambda_i=x_n\varepsilon_{ns}^m+y_n\varepsilon_{ns}^{unm}, \Delta_{ns} = \pi \rho_{ns}x_ny_n(\varepsilon_{ns}^m-\varepsilon_{ns}^{unm})^2, \varepsilon_{ns}^{m(unm)}=\varepsilon^{m(unm)}+(U^{m(unm)}/2)N_n^{m(unm)}-s(U^{m(unm)}/2)M_n^{m(unm)}, \varepsilon^{m(unm)}$ and $U^{m(unm)}$ are the nuclear potential energy and the electron-electron Coulomb interaction energy in magnetized (unmagnetized) Pt atoms, $N_n^{m(unm)}$ and $M_n^{m(unm)}$ are the number of outermost electrons and the atomic magnetic moment in the n-th layer of magnetized (unmagnetized) Pt atoms. $\rho_{ns}$ is the local density of states for electrons with spin s in the n-th layer.

\begin{table}[h]
\footnotesize
\caption{\label{tab:table1}%
Parameters of YIG/Pt/YIG used in the simulation.
}
\begin{ruledtabular}
\begin{tabular}{lcc}  
Parameters & Symbol & Value \\
\colrule
{Magnetized atoms' magnetic moment} & {$\mu_0$} & {0.2$\mu_B$ \cite{21}} \\ 
{Characteristic length of the MPE} & {$d_0$} & {0.4 nm \cite{21}} \\
{Gilbert damping constant of YIG} & {$\alpha$} & {0.001 \cite{32}} \\
{Electrons and magnons coupling} & {$\eta$} & {8 \cite{29}} \\
{On-site energy of YIG} & {$\varepsilon_i$} & {1 eV \cite{9,33,34}} \\
{Nearest neighbor transition energy} & {$t_{ij}$} & {-0.4 eV \cite{9,33,34}} \\
{Spin Hall angle of Pt} & {$\theta_{SH}$} & {0.01 \cite{30}} \\
{Conductivity of Pt} & {$\sigma$} & {$10^7$ $\Omega^{-1}m^{-1}$ \cite{35}} \\
{Spin diffusion length of Pt} & {$\lambda_{sd}$} & {1.5 nm \cite{36}} \\
{Length of Pt} & {$l$} & {100 $\mu$ m} \\
{Width of Pt} & {$w$} & {10 $\mu$ m} \\
{Thickness of Pt} & {$t$} & {2 nm} \\
{Real part of spin mixing conductance} & {$G_r$} & {$10^{15}$ $\Omega^{-1}m^{-2}$ \cite{28}} \\
\end{tabular}
\end{ruledtabular}
\end{table}

\begin{figure}[htbp]
\includegraphics[width = \linewidth]{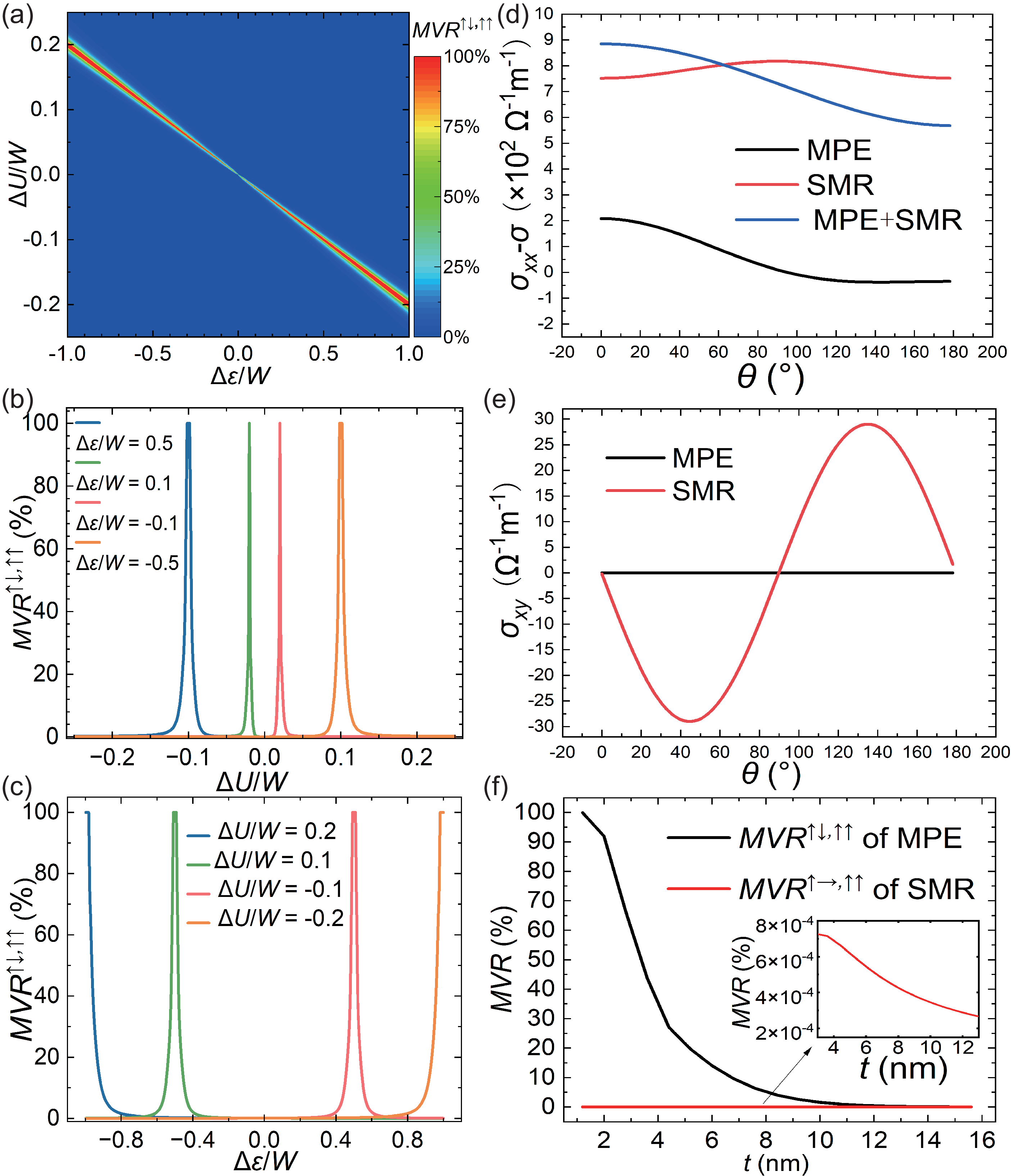}
\caption{(a) The dependence of $MVR^{\uparrow \uparrow, \uparrow \downarrow}$ on $\Delta \varepsilon$ and $\Delta U$. (b) The $\Delta U$ and (c) $\Delta \varepsilon$ dependence of MVR extracted from Fig. 2(a). The $\theta$ dependence of (d) longitudinal and (e) transverse conductance induced by MPE and SMR when $\Delta \varepsilon$/W = -0.98, $\Delta U$/W = -0.05. (f) The dependence of MVR on the thickness of Pt when $\Delta \varepsilon$/W = -0.528, $\Delta U$/W = 0.1035.}
\end{figure}

The dependence of $MVR^{\uparrow \uparrow, \uparrow \downarrow}$ on the difference of nuclear potential energy $\Delta \varepsilon$ and electron-electron Coulomb interaction energy $\Delta U$ between magnetized and non-magnetized Pt atoms is shown in Fig. 2(a). The parameters used in calculation are displayed in Table I. By choosing suitable nuclear potential energy and Coulomb interaction, the theoretical $MVR^{\uparrow \uparrow, \uparrow \downarrow}$ of the MV can reach 100\%, even comparable to the giant or tunneling magnetoresistance (GMR or TMR). We can see from Eq. (2) that $\Delta U/\Delta \varepsilon$ affects the $MVR^{\uparrow \uparrow, \uparrow \downarrow}$.  Because the number of electrons in the outermost layer of Pt is $N_{n}^{m}=N_{n}^{\text {unm }}=10$, when $\Delta U /\left(\Delta \varepsilon+N_{n}^{m} \Delta U / 2\right)$ is larger, the conductance difference between parallel and antiparallel state is larger. Therefore, the slope corresponding to the maximum $MVR^{\uparrow \uparrow, \uparrow \downarrow}$ is $\frac{\Delta U}{\Delta s}=-\frac{2}{N_{n}^{m}}=-0.2$, which is consistent with the calculation result shown by the oblique line in Fig. 2 (a) and peak in Fig. 2 (b, c). 
The MV also has potential to serve as a magnetic field sensor. 
We assume that both the applied current and the magnetization of the bottom YIG layer are along the x direction, and the magnetization of the top YIG layer deviate $\theta$ from x direction. When $\theta$ is changed, on the one hand, the magnetization of the magnetized Pt atoms adjacent to the top YIG-Pt interface are changed due to the MPE, and on the other hand, the conductance in Pt layer is changed due to the spin Hall magnetoresistance (SMR) effect \cite{28}, the $\theta$ dependence of longitudinal and transverse conductance is shown in Eq. (3) (see Supplementary Material for the detailed derivation).

The $\theta$ dependence of the longitudinal and transverse conductance is shown in Fig. 2 (d, e), the sum conductanceis $\sigma^{ {sum }}_{xx}=1 /\left(1 / \sigma^{MPE}_{xx}+1 / \sigma^{SMR}_{xx}-1 / \sigma\right)$. We can see that the longitudinal conductance $\sigma_{xx}$ remains the same for $\theta$ and 180°-$\theta$ states when only SMR is considered, but when MPE is present, we can distinguish the two states by only measuring the $\sigma_{xx}$. And these two states can also be distinguished by measuring transvers conductance $\sigma_{xy}$, as shown in Fig. 2(e). Finally, we calculate the $MVR^{\uparrow \uparrow, \uparrow \downarrow}$ induced by MPE and $MVR^{\uparrow \rightarrow, \uparrow \downarrow} \equiv\left(\sigma_{x x, \uparrow \rightarrow}-\sigma_{x x, \uparrow \downarrow}\right) / \sigma_{x x, \uparrow \rightarrow}$ induced by SMR as a function of the thickness of the Pt layer $t$ under the condition of $\Delta \varepsilon/W$ = -0.528 and $\Delta U/W$ = 0.1035, as shown in Fig. 2(f), where $\sigma_{x x, \uparrow \rightarrow}$ is longitudinal conductance when the magnetization of top YIG layer is along y direction and the magnetization of bottom YIG layer is along x direction, $W$ is the energy band width. The result demonstrates that as $t$ increases, the MVR exhibits an exponential decay. 

\begin{widetext}
\begin{equation}\label{eq3}
\begin{array}{ll}
\sigma_{x x}^{SMR}=&\sigma+\theta_{\mathrm{SH}}^{2}\sigma\left[ \frac{2 \lambda\tanh \frac{\mathrm{t}}{2 \lambda}}{\mathrm{t}}\right.+     (1+\frac{\left(\frac{2 \mathrm{e} \lambda}{\sigma \tanh \frac{\mathrm{t}}{\lambda}}+\frac{\operatorname{e \sigma \tanh} \frac{\mathrm{t}}{\lambda}+2 \mathrm{e} \lambda \mathrm{G}_{\mathrm{r}} \sin ^{2} \theta}{\sigma \mathrm{G}_{\mathrm{r}} \cos ^{2} \theta \tanh \frac{\mathrm{t}}{\lambda}}\right)\left(\frac{2 \mathrm{e} \lambda}{\sigma \tanh \frac{\mathrm{t}}{\lambda}}+\frac{2 \mathrm{e} \lambda}{\sigma \sinh \frac{\mathrm{t}}{\lambda}}+\frac{\mathrm{e}}{\mathrm{G}_{\mathrm{r}}}\right)}{\left(\frac{2 \mathrm{e} \lambda}{\sigma \sinh \frac{\mathrm{t}}{\lambda}}\right)^{2}-\left(\frac{2 \mathrm{e} \lambda}{\sigma \tanh \frac{\mathrm{t}}{\lambda}}+\frac{\mathrm{e}}{\mathrm{G}_{\mathrm{r}}}\right)\left(\frac{2 \mathrm{e} \lambda}{\sigma \tanh \frac{\mathrm{t}}{\lambda}}+\frac{e \sigma \tanh \frac{\mathrm{t}}{\lambda}+2 \mathrm{e} \lambda \mathrm{G}_{\mathrm{r}} \sin ^{2} \theta}{\sigma \mathrm{G}_{\mathrm{r}} \cos ^{2} \theta \tanh \frac{\mathrm{t}}{\lambda}}\right)})     \frac{\lambda\sinh \frac{\mathrm{t}}{\lambda}\tanh^2 \frac{\mathrm{t}}{2 \lambda}}{\mathrm{t}}\\
&+\frac{\frac{2 \mathrm{e} \lambda}{\sigma \tanh \frac{\mathrm{t}}{\lambda}}+\frac{2 \mathrm{e} \lambda}{\sigma \sinh \frac{\mathrm{t}}{\lambda}}+\frac{\mathrm{e}}{\mathrm{G}_{\mathrm{r}}}}{\left(\frac{2 \mathrm{e} \lambda}{\sigma \sinh \frac{\mathrm{t}}{\lambda}}\right)^{2}-\left(\frac{2 \mathrm{e} \lambda}{\sigma \tanh \frac{\mathrm{t}}{\lambda}}+\frac{\mathrm{e}}{\mathrm{G}_{\mathrm{r}}}\right)\left(\frac{2 \mathrm{e} \lambda}{\sigma \tanh \frac{\mathrm{t}}{\lambda}}+\frac{\operatorname{e} \sigma \tanh \frac{\mathrm{t}}{\lambda}+2 \mathrm{e} \lambda \mathrm{G}_{\mathrm{r}} \sin ^{2} \theta}{\sigma \mathrm{G}_{\mathrm{r}} \cos ^{2} \theta \tanh \frac{\mathrm{t}}{\lambda}}\right)}\left.              \frac{2 \mathrm{e} \lambda^2 \tanh ^2 \frac{\mathrm{t}}{2 \lambda}}{\sigma \mathrm{t}}\right]\\
\sigma_{x y}^{S M R}=&\theta_{\mathrm{SH}}^{2}\frac{\tan \theta (\frac{2 \mathrm{e} \lambda}{\sigma \tanh \frac{\mathrm{t}}{\lambda}}+\frac{2 \mathrm{e} \lambda}{\sigma \sinh \frac{\mathrm{t}}{\lambda}}+\frac{\mathrm{e}}{\mathrm{G}_{\mathrm{r}}})}{\left(\frac{2 \mathrm{e} \lambda}{\sigma \sinh \frac{\mathrm{t}}{\lambda}}\right)^{2}-\left(\frac{2 \mathrm{e} \lambda}{\sigma \tanh \frac{\mathrm{t}}{\lambda}}+\frac{\mathrm{e}}{\mathrm{G}_{\mathrm{r}}}\right)\left(\frac{2 \mathrm{e} \lambda}{\sigma \tanh \frac{\mathrm{t}}{\lambda}}+\frac{\operatorname{e} \sigma \tanh \frac{\mathrm{t}}{\lambda}+2 \mathrm{e} \lambda \mathrm{G}_{\mathrm{r}} \sin ^{2} \theta}{\sigma \mathrm{G}_{\mathrm{r}} \cos ^{2} \theta \tanh \frac{\mathrm{t}}{\lambda}}\right)}           \frac{2 \mathrm{e} \lambda ^2 \tanh ^2 \frac{\mathrm{t}}{2 \lambda} }{\mathrm{t}}
\end{array}
\end{equation}
\end{widetext}

\begin{figure}[htbp]
    \includegraphics[width = \linewidth]{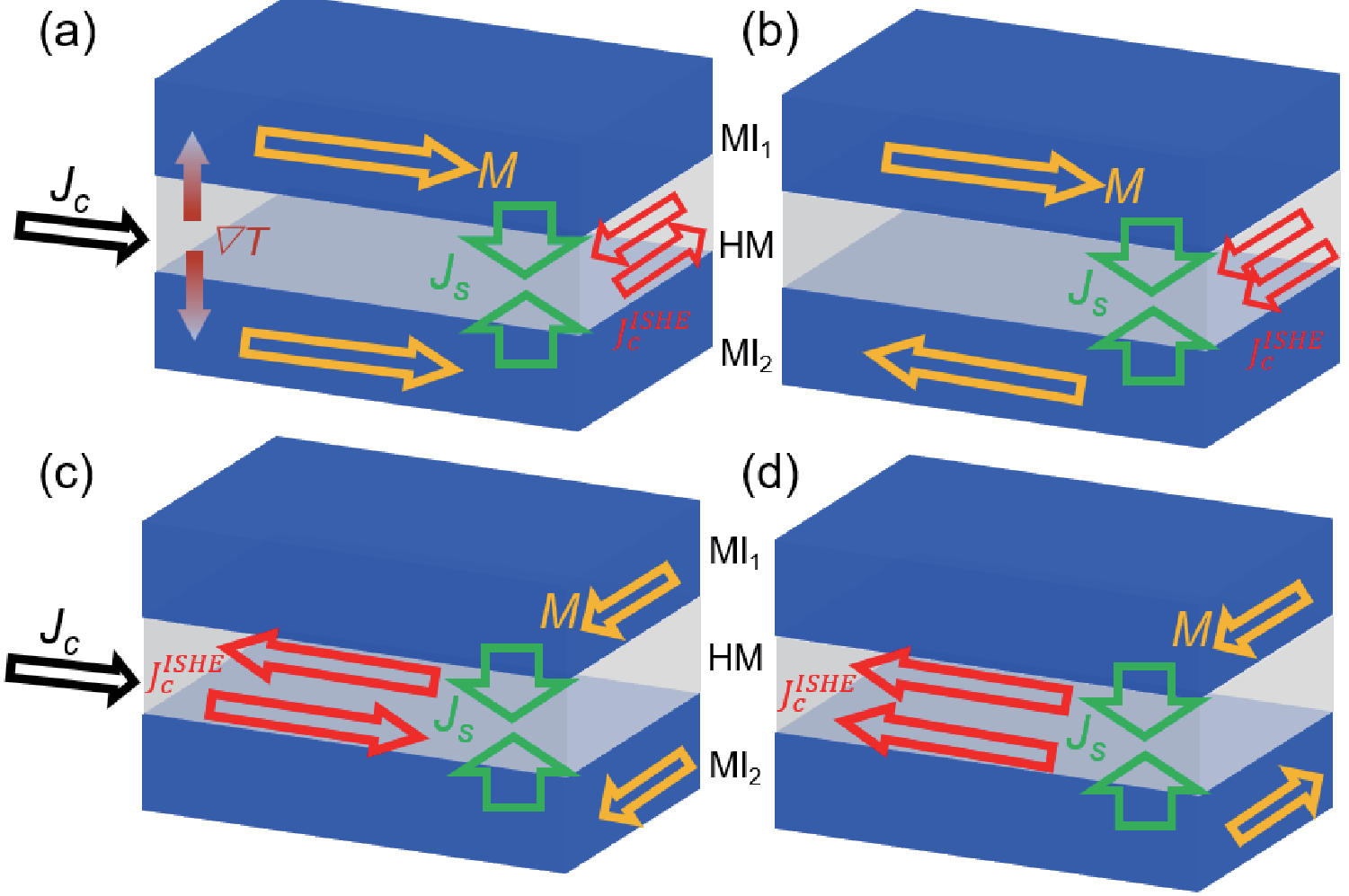}
    \caption{Schematic of the influence of Joule heating on the measurement signal based on the SSE. In the first configuration, an electron current is aligned with the magnetization of the MI layer. This results in different transverse voltage between the (a) parallel and (b) antiparallel state MV. In the second configuration, an electron current is applied perpendicular to the magnetization of the MI layer. This results in different longitudinal resistance between the (c) parallel and (d) antiparallel state MV.}
\end{figure}

The Joule heating induced by the electron current in the Pt layer will create a temperature gradient between the Pt layer and YIG layer. The temperature gradient, in turn, induces a magnon current that permeates the Pt layer, consequently influencing electron transport in the Pt layer due to ISHE. More specifically, When the current direction is aligned with the magnetization of the YIG layer, as depicted in Fig. 3(a, b).
According to the relationship between the ISHE current and the spin current $\mathbf{j}_{\mathbf{c}}^{\text {ISHE }} \propto \mathbf{j}_{s} \times \sigma$, we can deduce that the spin current will give rise to an transverse ISHE voltage. 
On the other hand, when the direction of the electron current and magnetization are perpendicular, as depicted in Fig. 3(c, d).
Employing a similar analysis to the previous scenario, we can conclude that a temperature gradient will 
influence the longitudinal resistance.

To further quantify the effect of temperature gradient on the electrical transport properties of the Pt layer, we set the upper and lower YIG layers with temperature $T_1$, and a middle Pt layer with temperature $T_2$, where $T_2$ $>$ $T_1$. Considering the nearest neighbor Heisenberg exchange interaction, the Hamiltonian describing the magnon system in the top YIG layer interface is
\begin{equation}\label{eq4}
\hat{H}_{YIG}=\sum_{i}{\varepsilon_i \hat{a}_i^\dag \hat{a}_i}+\sum_{<i,j>}{t_{ij}\hat{a}_i^\dag \hat{a}_j}
\end{equation}

Where $\varepsilon_i$ is on-site energy, $t_{ij}$is the nearest neighbor transition energy, $<i,j>$ represents the sum over the neighboring lattice sites. Then we can use non equilibrium Green’s method to calculate the magnon current injected from YIG layers to Pt layer \cite{17,29} (see Supplementary Material for the detailed derivation).

The inverse spin Hall voltage along the y direction is calculated according to the formula.

\begin{equation}\label{eq5}
V_{ISHE}^y=\frac{\rho L}{tw}\int{\mathbf{j_c^{ISHE}}\cdot\mathbf{e_y}dS}
\end{equation}
where $\rho$ is the resistivity of Pt layer, $L$, $t$, $w$ is the length, thickness and width of the Pt layer respectively. $j_{c}^{ISH E}=\frac{2 e \theta_{\mathrm{SH}} \lambda_{\mathrm{sd}}}{\hbar t} \tanh \left(\frac{t}{2 \lambda_{\mathrm{sd}}}\right) j_{m}$ \cite{30,31} is the inverse spin Hall electric current, where $\theta_{SH}$ is spin Hall angle of Pt, $\lambda_{sd}$ is spin diffusion length of Pt, $j_m$ is magnon current injected from YIG layers. The parameters used in simulation are displayed in Table I.

\begin{figure}[htbp]
\includegraphics[width = \linewidth]{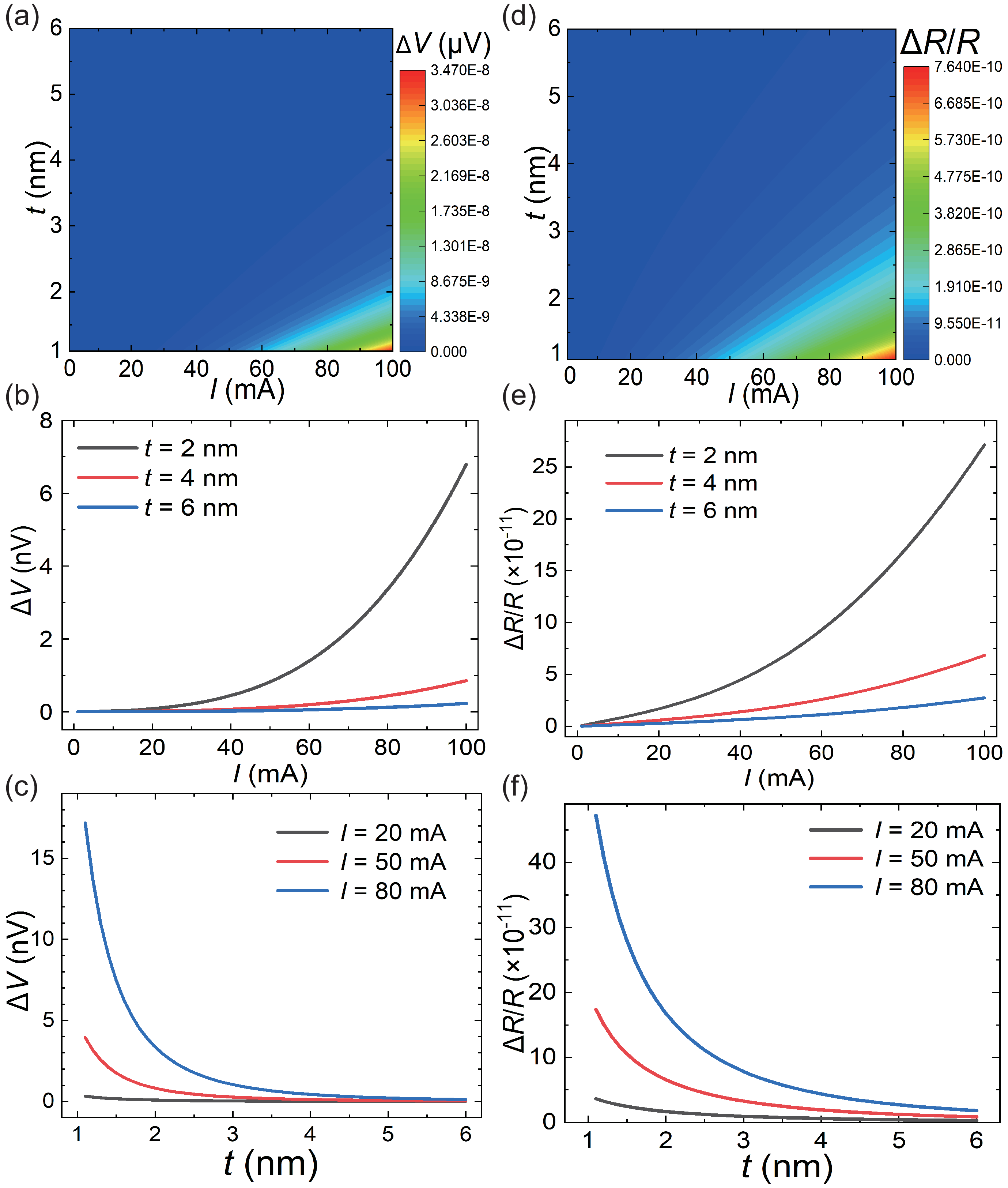}
\caption{ (a) The dependence of $\Delta V$ on $I$ and $t$ in the first configuration. The (b) $I$ and (c) $t$ dependence of $\Delta V$ extracted from Fig. 4(a). (d) The dependence of $\Delta R/R$ on the $I$ and $t$ in the second configuration. The (e) $I$ and (f) $t$ dependence of $\Delta R/R$ extracted from Fig. 4(d).}
\end{figure}

Here we calculate the relationship between current $I$, Pt layer thickness $t$, and the transverse voltage difference $\Delta V$ between parallel and antiparallel state, depicted in Fig. 4(a). The ambient temperature is 300 K, the results at different ambient temperatures is also calculated 
(see Supplementary Material for the detail). 
As $I$ increases and the $t$ decreases, the temperature gradient generated by the current in the Pt layer increases, therefore $\Delta V$ generated by the SSE increases. We extract the relationship between $I$, $t$ and $\Delta V$, which is shown in Fig. 4(b, c). We can see that when $I$ reaches 100 mA, it is feasible to generate a measurable voltage up to the magnitude of nV. In the second configuration, we calculate the relationship between longitudinal resistance change rate $\Delta R/R$ and $I$, $t$, as shown in Fig. 4(d). 
We also extract the relationship between the $\Delta R/R$ and $I$, $t$ and showed in Fig. 4(e, f). $\Delta R/R$ increases as $I$ increases and $t$ decreases, but the relative change is small (on the order of $10^{-11}$). Therefore, measuring $\Delta R/R$ in experiments poses considerable challenges.
Fig. 4(b, c, e, f) demonstrates that both $\Delta V$ and $\Delta R/R$ display an approximat quadratic relationship with respect to $I$ and are inversely proportional to $t$. This can be attributed to the direct proportionality of the heat produced by $I^2$, thereby generating a temperature gradient that is also proportional to $I^2$ and inversely proportional to $t$. Furthermore, the signal induced by the SSE is also proportional to $I^2$ and inversely proportional to $t$.

In this work, we propose the IMVE based on the MPE. We conduct a theoretical investigation into the dependence of $MVR^{\uparrow \uparrow, \uparrow \downarrow}$ on the difference of electron structure between magnetized and non-magnetized Pt atoms. It is found that $MVR^{\uparrow \uparrow, \uparrow \downarrow}$ can reach 100\% theoretically. Furthermore, we study the relationship between $MVR^{\uparrow \uparrow, \uparrow \downarrow}$ and the thickness of the Pt layer, revealing an exponential decay as the thickness increases. The dependence of Pt layer conductance on the relative angle between the magnetizations of two MI layers is also calculated, illustrating the potential of MV as a magneto-sensitive magnonic sensor. The influence of Joule heating on the measurement signal based on the SSE is also explored. Two designed configurations are proposed according to whether the electron current is parallel or perpendicular to the magnetization of the MI layer. In the parallel configuration, the transverse voltage differs between the parallel and antiparallel MV states. While in the perpendicular configuration, the longitudinal resistance differs. Quantitative numerical results indicate that it is feasible to measure the voltage signal using the first configuration. Our work contributes valuable insights for the design, development and integration of magnon devices.

This work is financially supported by the National Key Research and Development Program of China (MOST) (Grants No. 2022YFA1402800), the National Natural Science Foundation of China (NSFC) (Grants No. 51831012, 12134017, 11974398), and partially supported by the Strategic Priority Research Program (B) [Grant No. XDB33000000, Youth Innovation Promotion Association of CAS (2020008)].
\nocite{*}

\providecommand{\noopsort}[1]{}\providecommand{\singleletter}[1]{#1}%

\end{document}